\documentclass[review, 10pt]{elsarticle}

\usepackage{graphicx}
\usepackage{amsmath,amsfonts,amsthm,hyperref}
\usepackage{booktabs}
\usepackage[justification=centering]{caption}
\usepackage{subcaption}
\usepackage{multirow}
\usepackage{xcolor}
\usepackage[flushleft]{threeparttable}
\usepackage{tabulary}
\usepackage{longtable}
\usepackage{lscape}
\usepackage[pass]{geometry}
\usepackage{makecell}

\journal{Biomedical Signal Processing and Control}

\biboptions{sort&compress}

\begin{document}

\begin{frontmatter}

\title{CardioPHON: Quality assessment and self-supervised pretraining for screening of cardiac function based on phonocardiogram recordings}

\author[mymainaddress]{Vladimir Despotovic\corref{mycorrespondingauthor}}
\cortext[mycorrespondingauthor]{Corresponding author}
\ead{vladimir.despotovic@lih.lu}
\address[mymainaddress]{Bioinformatics and AU Unit, Department of Medical Informatics, Luxembourg Institute of Health, Luxembourg}

\author[mysecondaryaddress]{Peter Pocta}
\address[mysecondaryaddress]{Department of Multimedia and Information-Communication Technology, Faculty of Electrical Engineering and Information Technology, University of Zilina, Slovakia}
\author[myternaryaddress]{Andrej Zgank}
\address[myternaryaddress]{Laboratory for Digital Signal Processing, Faculty of Electrical Engineering and Computer Science, University of Maribor, Slovenia}

\begin{abstract}
Remote monitoring of cardiovascular diseases plays an essential role in early detection of abnormal cardiac function, enabling timely intervention, improved preventive care, and personalized patient treatment. Abnormalities in the heart sounds can be detected automatically via computer-assisted decision support systems, and used as the first-line screening tool for detection of cardiovascular problems, or for monitoring the effects of treatments and interventions. 
We propose in this paper CardioPHON, an integrated heart sound quality assessment and classification tool that can be used for screening of abnormal cardiac function from phonocardiogram recordings. The model is pretrained in a self-supervised fashion on a collection of six small- and mid-sized heart sound datasets, enables automatic removal of low quality recordings to ensure that subtle sounds of heart abnormalities are not misdiagnosed, and provides a state-of-the-art performance for the heart sound classification task. 
The multimodal model that combines audio and socio-demographic features demonstrated superior performance, achieving the best ranking on the official leaderboard of the 2022 George B. Moody PhysioNet heart sound challenge, whereas the unimodal model, that is based only on phonocardiogram recordings, holds the first position among the unimodal approaches (a total rank 4), surpassing the models utilizing multiple modalities. CardioPHON is the first publicly released pretrained model in the domain of heart sound recordings, facilitating the development of data-efficient artificial intelligence models that can generalize to various downstream tasks in cardiovascular diagnostics.
\end{abstract}

\begin{keyword}
heart sound, phonocardiogram, quality assessment, self-supervised learning, cardiac function. 
\end{keyword}

\end{frontmatter}


\section{Introduction} 
Cardiovascular diseases (CVD) represent the leading cause of mortality worldwide, covering approximately one-third of deaths globally~\cite{chen2021deep}, and 300,000 deaths in North America annually~\cite{chan2019contactless}.  
Cardiac auscultation is a simple and non-invasive, yet useful first-line CVD screening method, that enables detection of the structural heart abnormalities or assessing them in a differential diagnosis. However, it requires substantial clinical experience and skilled physicians \cite{Mangione97}. Moreover, some heart sounds fall within the frequency range inaudible for humans~\cite{chen2021deep}. Computer-assisted decision support systems based on auscultation can be used to assist timely diagnosis and treatment of CVD patients~\cite{clifford2017recent, reyna2022heart}. 

The heart sounds are recorded during cardiac auscultation as phonocardiogram (PCG) signals, using digital stethoscopes that capture the acoustic vibrations made by the heart with a microphone-transducer. With the introduction of wearable PCG sensors \cite{Lin21} that have a tremendous potential for in-home cardiac care, real-time remote CVD screening has become feasible. An in-home or AAL environment~\cite{despotovic2022audio} is more challenging than a clinical environment for heart sounds' acquisition, given that unskilled/naive users mostly operate the PCG sensors. A heart sound captured in a clinical environment is already contaminated by internal physiological body noises and ambient artifacts \cite{Altuve2020}, and even more noisy when recorded in real-world scenarios. Therefore, assessing the recording quality is essential to ensure the robustness, reliability and clinical interpretability of the models for screening of CVDs. 

The heart sound quality assessment was mostly implemented using the traditional machine learning algorithms, such as logistic regression to classify a signal quality of PCG recordings acquired by the non-experts using commercial medical-grade and mobile phone-based electronic stethoscopes~\cite{springer2016automated}, Random Forest trained with sensor-agnostic spectral domain features~\cite{das2017novel}, or Support Vector Machine (SVM) for binary quality classification (unacceptable vs. acceptable), and with three quality levels (unacceptable, good and excellent)~\cite{tang2021automated}. An approach for automatic assessment of heart and breathing signal quality from noisy neonatal chest sounds was designed in~\cite{grooby2021neonatal}, and further extended for the real-time multi-level neonatal heart and lung sound quality assessment in~\cite{grooby2022real}. Integrated heart sound quality assessment and classification approach was proposed in~\cite{zabihi2016heart}, being ranked second in the PhysioNet/Computing in Cardiology Challenge 2016. Another approach uses SVM for assessing the quality of PCG recordings and localization of heart sounds, by minimizing the inter- and intra-class variations between S1 and S2 sounds~\cite{akram2018analysis}, whereas heart sound classification in~\cite{mei2021classification} uses wavelet scattering transform. 

Recent heart sound classification approaches used an ensemble of AdaBoost and Convolutional Neural Network (CNN)~\cite{potes2016ensemble}, or Recursive Neural Network (RNN) to produce observations for Hidden semi-Markov Models~\cite{mcdonald2022detection}. Ballas et al.~\cite{ballas2022listen2yourheart} proposed the first heart sound classification approach employing self-supervised learning, but with a moderate performance, ranked $13^{th}$ for the murmur classification and $7^{th}$ for the outcome classification at the George B. Moody PhysioNet Challenge 2022~\cite{reyna2022heart, Oliveira22} (PhysioNet 2022 in further text), most likely due to the limited amount of PCG recordings used for model pretraining. The authors in~\cite{kamson2024exploring, panah2023exploring} deployed pretrained wave2vec 2.0~\cite{Baevski20} for the murmur classification task, i.e., the second task of the George B. Moody PhysioNet Challenge 2022~\cite{reyna2022heart, Oliveira22}, also involving the limited amount of PCG recordings for model pretraining. More recently, a self-supervised learning approach entitled Masked Modeling Duo, specifically designed for the murmur classification task, was introduced in~\cite{niizumi2024exploring}, again using a limited amount of heart sound data. 

Access to large-scale data is a critical aspect of every machine-learning task. The latest development in speech and language technologies, i.e., the large language models used in natural language processing (e.g., GPT4, LLaMA2) or automatic speech recognition (e.g., Whisper \cite{radford2023robust}, wav2vec 2.0 \cite{Baevski20}), showed remarkable improvements due to access to large quantities of training data. Unfortunately, there are no heart sound datasets available on a similar scale. To circumvent this limitation we have identified all small and medium-sized heart sound datasets publicly available to date, and used them either to pretrain a deep learning model in a self-supervised way that does not require any annotation, or to fine-tune the model pretrained with the large-scale general purpose audio data. 

Our contribution in this paper is three-fold: 

\begin{enumerate}
\item{We developed a model for quality assessment of PCG recordings inspired by the features and quality annotations proposed in~\cite{tang2021automated, grooby2021neonatal}. It was used to identify the heart sounds that did not satisfy the minimum quality required for further processing, and prepare the data for the subsequent heart sound classification task.}
\item{We pretrained the model for heart sound classification in a self-supervised manner using six publicly available heart sound datasets, making it the largest and the most diverse collection of PCG recordings so far, and enabling the models to capture sufficient domain-specific features. To the best of our knowledge, this is the first released pretrained model in the domain of heart sound recordings. It may facilitate future development of data-efficient artificial intelligence models that can transfer well to various downstream tasks in cardiovascular diagnostics.}
\item{We integrated the quality assessment into the heart sound classification and named the integrated model CardioPHON. The model was evaluated on the held-out George B. Moody PhysioNet Challenge 2022 dataset~\cite{reyna2022heart, Oliveira22}, a widely adopted benchmark dataset for heart sounds, achieving state-of-the-art results. }
\end{enumerate}

\section{Material and methods}  
\label{sec:methods}

\subsection{Datasets}
\label{subsec:datasets}

To create the basis for pretraining models in a self-supervised fashion, we have identified major publicly available datasets of heart sounds, collected both in clinical and non-clinical environments. The following data selection criteria were applied: 1) The availability of data in the open access mode or on request; 2) The total length of PCG recordings longer than several minutes; 3) More than one participant included in the dataset. The following datasets were included in the experiments (see Table~\ref{tab:datasets}):

\begin{table*}[b]
   \caption{Heart sound datasets and their characteristics.}
   \label{tab:datasets}
   \scriptsize
   \centering
   \hspace*{-3.25cm}\begin{tabular}{lccccccc}
     \toprule
     & \textbf{Pascal} & \textbf{HSCT11} & \textbf{CDHS} & \textbf{Ephnogram} & \textbf{Open Heart} & \textbf{CinC16} & \textbf{CirCor22} \\
     \midrule
      \textbf{Participants} & NA & 206 & 76 & 24 & NA & 1,072 & 1,568 \\
     \textbf{Recordings} & 832 & 412 & 3,875 & 69 & 1,000 & 3,240 & 5,272 \\
     \textbf{File length [s]} & 0.8 -- 27.9 & 17.9 -- 71.1 & 15.0 -- 34.1 & 1,800 & 1.1 -- 4.0 & 5.3 -- 122.0 & 4.8 -- 80.4 \\
     \textbf{Duration [h]} & 1.6 & 5.2 & 18.1 & 30.6 & 0.7 & 20.2 & 33.5 \\
     \textbf{Age categories} & mixed & adults & adults & adults & NA & mixed & children \\
     \textbf{Additional information} & segment. & socio-dem. & quality score & ECG, socio-dem. & NA & segment. & segment., socio-dem. \\
     \textbf{Sampling frequency [kHz]} & 4 / 44.1 & 11 & 1 & 8 & 8 & 2 & 4 \\
     \textbf{Quality assessment}  & yes & yes & yes & no & no & yes & no \\
     \textbf{Dataset usage}  & Pretraining & Pretraining & Pretraining & Pretraining & Pretraining & Pretraining & Evaluation \\
     \bottomrule
   \end{tabular}
\end{table*}

\begin{enumerate}
\item \textbf{Pascal} dataset \cite{Gomes2013} was a part of the PASCAL 2011 Heart Sounds Challenge, and contains PCG recordings acquired in clinical and non-clinical environments, either by a digital stethoscope, or via a mobile phone. The dataset was used for the heart sound segmentation task, to identify S1 and S2 heart sounds, and for the heart sound classification task, to distinguish between the normal, murmur, extra heart sound, and artifacts.

\item \textbf{Heart Sounds Catania 2011} (HSCT11) \cite{Spadaccini13} consists of PCG recordings collected from multiple auscultation locations (mitral, pulmonary, aortic, tricuspid), recorded using a digital electronic stethoscope. It contains no information about the health condition of the participants.

\item \textbf{Cardiac Disease Heart Sound} (CDHS) is a heart sound dataset~\cite{tang2021automated} collected in a clinical environment for pulmonary hypertension monitoring. CDHS was combined with 3 existing datasets – HSCT11, Pascal, and CinC16, and further underwent manual quality assessment labeling~\cite{tang2021automated}. 

\item \textbf{Ephnogram} \cite{Kazemnejad2021} contains simultaneous PCG and electrocardiogram (ECG) recordings, acquired from young adults (23-29 years) under rest and physical activity (walking, running and biking). ECG signal was excluded from our experiments. 

\item \textbf{Open Heart} is a collection of PCG recordings from multiple books and internet sources~\cite{yaseen2018classification}. Each recording contains exactly three heartbeats (2.44 seconds average length), and is classified into 5 categories (normal, aortic stenosis, mitral stenosis, mitral regurgitation, mitral valve prolapse).  

\item \textbf{CinC16} \cite{Liu2016} was used in the PhysioNet/Computing in Cardiology Challenge 2016 and contains PCG recordings from 9 independent databases, collected at different sampling frequencies with multiple devices, and downsampled to 2 kHz. The recordings were acquired from 4 different locations (aortic, pulmonary, tricuspid, and mitral) from both healthy subjects and patients with different CVDs. The dataset is divided into training, validation and test subsets; however, the test subset is kept private for official challenge evaluation and unavailable publicly.

\item \textbf{CirCor22 DigiScope} (CirCor22) \cite{Oliveira22} is the largest publicly available dataset used as a part of the PhysioNet 2022 challenge. It contains PCG recordings from 1,568 Brazilian children or adolescents, acquired with a digital stethoscope from 4 locations (aortic, pulmonary, tricuspid, and mitral). It contains 3,163 PCG recordings in the training and 2,109 in the validation and test sets, but also socio-demographic data, such as age group~\cite{williams2012standard}, sex, height, weight, and pregnancy status.

\end{enumerate}

\subsection{Quality assessment of PCG recordings}
\label{subsec:quality}
Pascal, HSCT11, CinC16, and CDHS datasets~\cite{tang2021automated} containing the quality scores ranging from 1 to 5 were used for development of the quality assessment model. A mapping of the quality scores to binary classification categories (“unacceptable” and “acceptable”), has followed the approach deployed in  ~\cite{tang2021automated}, i.e., the quality scores of 1, 2 and 3 were mapped to an unacceptable quality label, and the remaining ones, i.e., 4 and 5, to an acceptable quality label.

The auscultation theory states~\cite{mcgee2018auscultation} that a short screening time provides insufficient information to evaluate heart conditions. The file length analysis in Table~\ref{tab:datasets} shows that this was generally the case for the Open Heart dataset. We still decided to keep it, as it represents a non-negligible additional source of information for pretraining the model in the domain of heart sounds. Since the feature extraction for quality assessment was unstable for such short recordings, copies of individual PCG recording were concatenated to a minimum length of 6s before further processing. The same approach was applied to the short recordings in the Pascal dataset. Another data preprocessing step was needed for the Ephnogram dataset, where the original PCG recordings were 1,800s long. They were cut into 3,668 files of 30 seconds length, with the goal to establish similar conditions to other datasets and enable stable feature extraction.

\begin{figure}[t]
 \centering
    \begin{center}
        \includegraphics[width=8cm]{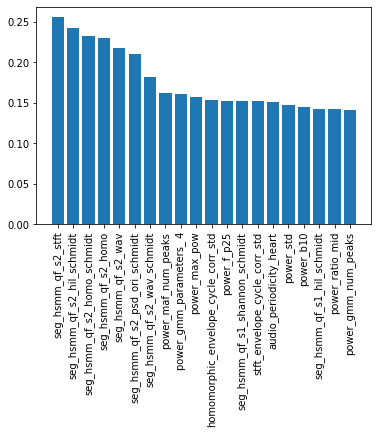}
    \end{center}
    \caption{Feature importance for the heart sound  quality assessment based on the mutual information criterion (20 most important features).}
    \label{fig:feat_importance}
\end{figure}

For PCG recordings available in Pascal, HSCT11, CinC16, and CDHS datasets, we followed the feature extraction approach described in~\cite{grooby2021neonatal}. The extracted features were ranked by their importance using the mutual information criterion, leaving only a subset composed of 20\% of the most relevant features (out of 416 features in total) for building the quality assessment model. Twenty most important features are shown in Figure~\ref{fig:feat_importance}. Mutual information measures the dependency between the independent variable (feature) and the dependent variable (quality label), and represents a non-negative number, with zero denoting independence between two variables, and higher values denoting bigger dependency \cite{Ross2014}.

Pascal, HSCT11, CinC16, and CDHS datasets were divided randomly into training and test subsets containing 80\% and 20\% of data, respectively. Stratified sampling was done to preserve approximately equal quality label distribution in the training and test subsets. The model designed for quality assessment was trained using the Support Vector Machines (SVM), Random Forest (RF) and Gradient Boosting (GB) machine learning models, as well as the Voting Classifier (VC), that combined all of them together, and selected the quality label with the biggest average predicted probability (soft voting criterion). 

Once the model for quality assessment was trained, it was used to evaluate the quality labels for all the other datasets listed in Table~\ref{tab:datasets}. Note that the quality assessment model was developed for the sampling frequency of 1 kHz, as all labeled datasets used for model design were resampled to 1 kHz~\cite{tang2021automated}.

\subsection{Heart sound classification from PCG recordings}
\label{subsec:heart_classification}

Heart sound data, like many other medical data, suffer from a scarcity of labeled examples due to the expertise required for accurate labeling. To mitigate this we propose a self-supervised learning (SSL) approach that leverages unlabeled data to learn general heart sound representations, and provide a good initialization for supervised learning models, leading to faster convergence and improved accuracy. SSL requires defining a special pretext task that mimics a supervised learning problem for the unlabeled data \cite{Liu2022}. These tasks are designed so that solving them requires the model to learn meaningful representations of the data, without the need for human annotations. We pretrain the models using SSL on the collection of 6 small- to mid-scale unlabelled PCG datasets, and further use several transfer learning strategies for the downstream heart sound classification task.

\begin{figure*}[t]
 \centering
    \begin{center}
        \includegraphics[width=12cm]{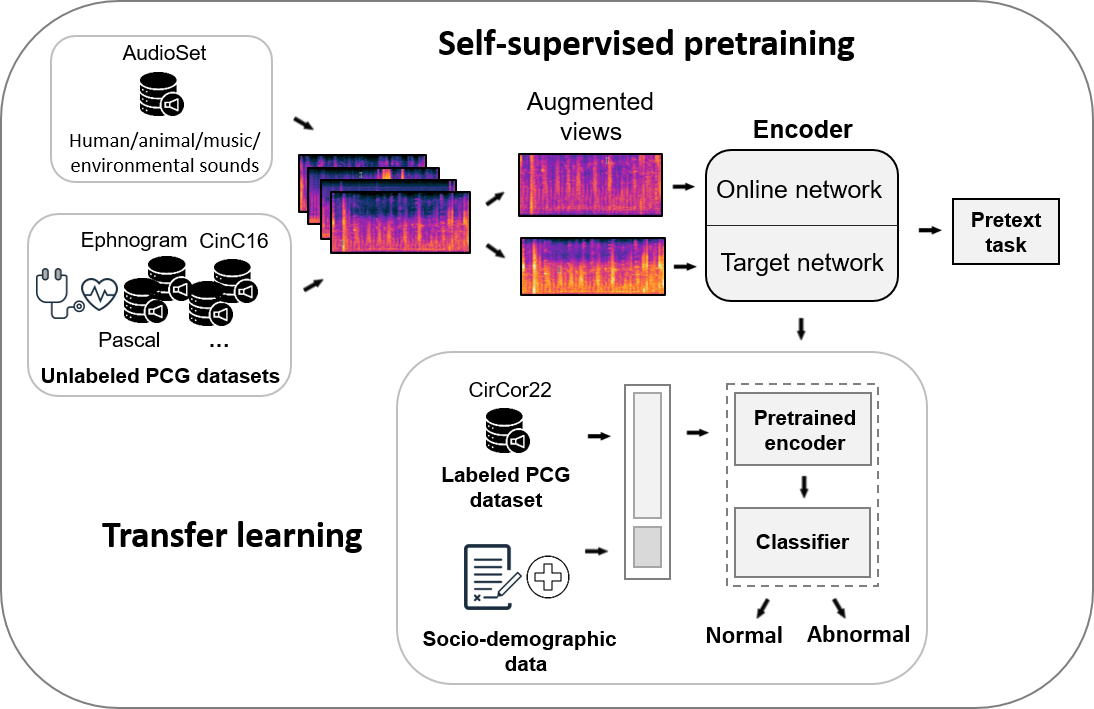}
    \end{center}
    \caption{Workflow of the transfer learning approach with self-supervised pretraining. In feature extraction (zero-shot learning) the representations are learned with the model pretrained on AudioSet only. In training from scratch the model is pretrained on a collection of 6 PCG datasets only. In fine-tuning the model pretrained on AudioSet is further retrained using 6 PCG datasets to adapt to a target domain. }
    \label{fig:workflow}
\end{figure*}

We use a method for self-supervised pretraining named Bootstrap Your Own Latent for Audio (BYOL-A)~\cite{niizumi2021byol-a}, which takes normalized 96x64 bin log-mel spectrograms as inputs, and creates two augmented versions of each spectrogram by shifting pitch and stretching time. They are further processed by two parallel networks, i.e., online and target networks, where the online network predicts the output representation of the target network, which is iteratively updated as the exponential moving average of the parameters of the online network \cite{Grill20}. We utilized the latest BYOL-A version, that combines local and global features while preserving frequency- and channel-wise information, and outputs 3,072-dimensional feature vectors \cite{niizumi2023byol-a}. The choice of BYOL-A for SSL was driven by the fact that the representations learned by BYOL-A are versatile and transferable across different audio-related tasks, making it a robust choice for general audio pretraining \cite{niizumi2021byol-a, niizumi2023byol-a}.

We analysed three different setups for the downstream heart sound classification task that leveraged transfer learning, i.e., feature extraction (zero-shot learning), training from scratch, and fine-tuning, as shown in Figure~\ref{fig:workflow}. 

\subsubsection{Feature extraction (zero-shot learning)}
\label{subsubsec:feat_extract}

In feature extraction, the base BYOL-A model was pretrained on the AudioSet, a general purpose large-scale dataset composed of 632 audio event classes (human/animal sounds, musical instruments, environmental sounds, etc.), that contains over 2 million human-labeled sound clips from YouTube videos \cite{Gemmeke2017}. A classification head is added on top of the base BYOL-A model and trained to re-purpose the previously learned feature maps, and match the domain and classes of the downstream heart sound classification task. On the other hand, weights of the pretrained BYOL-A model were kept frozen, i.e., zero-shot learning setup was applied. The rationale behind this was that low-level audio features are shared across different domains, i.e., the domains of general audio and heart sounds. The classification head was composed of two hidden layers with 256 neurons, with rectified linear unit (ReLU) activation function, dropout rate of 0.5, and an output layer with the number of neurons equal to the number of classes.

The training was done on the merged training and validation CirCor22 subsets, and a developed model was evaluated on the official CirCor22 test subset, which was made available to us by courtesy of the challenge organizers and the CirCor22 dataset creators. The model was trained using a single GPU (NVIDIA Quadro RTX 6000) for 20 epochs with the learning rate of $10^{-4}$, Adam optimizer, step learning rate scheduler that decays the learning rate by a factor of 10 after each 5 epochs, mini-batch size of 32, and cross-entropy loss function, as shown in Table~\ref{tab:hyperparameters}. We used Python (version 3.9.17) and PyTorch (version 2.0.0, CUDA 11.7) for all experiments and analyses in this study.

\begin{table*}[t]
   \caption{Hyperparameters of the classifier}
   \label{tab:hyperparameters}
   \scriptsize
   \centering
   \hspace*{-1cm}\begin{tabular}{cccccc}
     \toprule
      \textbf{Neurons/layer} & \textbf{Epochs} & \textbf{Batch size} & \textbf{Dropout rate} & \textbf{Learning rate} & \textbf{Learning rate decay} \\
     \midrule
     256 & 20 & 32 & 0.5 & $10^{-4}$ & 5 epochs* \\
     \bottomrule
   \end{tabular}
   \vskip 3mm   
    *Step learning rate decay reduced by a factor of 10 after each $n$ epochs.\\
\end{table*} 

\subsubsection{Training from scratch}
\label{subsubsec:scratch}

The model with the same architecture as the base BYOL-A, but with randomly initialized weights (i.e., it does not leverage pretraining on AudioSet), is trained from scratch on the collection of 6 publicly available PCG datasets (Pascal, HSCT11, CDHS, Ephnogram, Open Heart and CinC16) with 13,026 PCG recordings in total. CirCor22 dataset was used for an external evaluation, and was therefore not included in the pretraining data collection, as shown in Table~\ref{tab:datasets}. The rationale behind this setup was to investigate whether a training from scratch might be preferred over the feature extraction (zero-shot learning), given the mismatch between the domains of general audio and heart sounds. Evaluation was performed with the same classification head and model parameters as in the case of the feature extraction setup (see Table~\ref{tab:hyperparameters}). 

\subsubsection{Fine-tuning}
\label{subsubsec:finetuning}

Fine-tuning is combining the previous two approaches, i.e., we unfroze all the layers of the base BYOL-A model pretrained on the AudioSet dataset, and retrained it for 30 epochs using the collection of six publicly available PCG datasets (Pascal, HSCT11, CDHS, Ephnogram, Open Heart and CinC16), thus, customizing the base BYOL-A model to a new domain of the PCG recordings. The same pretraining parameters were set as in training from scratch, and evaluation
was performed with the same classification head and model parameters (see Table~\ref{tab:hyperparameters}). The rationale behind this setup was to evaluate whether the domain adaptation of the general audio to the domain of heart sounds may additionally improve the performance.

\subsubsection{Multimodal learning}
\label{subsubsec:multimodal}

Given the fact that CirCor22 dataset contains, beside PCG recordings, additional socio-demographic information for each participant in the study, such as age, gender, height, weight and pregnancy status, we ran two separate analyses for the PhysioNet 2022 challenge, i.e., one based only on features extracted from audio, and another fusing audio with socio-demographic variables. We wanted to evaluate whether socio-demographic variables improve the predictive capacity of audio features. The socio-demographic and audio features were fused using the feature-level fusion, by concatenating the feature vectors.

\begin{table*}[htbp]
   \caption{Study population characteristics of the CirCor22 dataset}
   \label{tab:demographic}
   \scriptsize
   \centering
   \begin{tabular}{llcccccc}
     \toprule
     & & \multicolumn{2}{c}{\textbf{Normal}} & \multicolumn{2}{c}{\textbf{Abnormal}} & \multicolumn{2}{c}{\textbf{Total}}\\
     \midrule
     \multicolumn{2}{l}{\textbf{Participants}}  & \multicolumn{2}{c}{813} & \multicolumn{2}{c}{755} & \multicolumn{2}{c}{1,568}\\
     \midrule
     \multirow{2}{*}{\textbf{Gender}} & \textbf{Female} & 421 & (51.8\%) & 360 & (47.7\%) & 781 & (49.8\%)\\
     & \textbf{Male} & 392 & (48.2\%) & 395 & (52.3\%) & 787 & (50.2\%)\\
     \midrule
     \multirow{5}{*}{\textbf{Age}} & \textbf{Neonate} & 5 & (0.6\%) & 5 & (0.7\%) & 10 & (0.6\%)\\
     & \textbf{Infant} & 73 & (9.0\%) & 119 & (5.8\%)& 192 & (12.2\%)\\
     & \textbf{Child} & 576 & (70.9\%) & 534 & (70.7\%) & 1,110 & (70.8\%) \\
     & \textbf{Adolescent} & 58 & (7.1\%) & 75 & (9.9\%) & 133 & (8.5\%) \\
     & \textbf{None} & 101 & (12.4\%) & 22 & (2.9\%) & 123 & (7.9\%) \\
     \midrule
     \multirow{5}{*}{\textbf{Age-corrected BMI}} & \textbf{Underweight} & 30 & (3.7\%) & 58 & (7.7\%) & 88 & (5.6\%) \\
     & \textbf{Normal weight} & 407 & (50.1\%) & 400 & (53.0\%) & 807 & (51.5\%) \\
     & \textbf{Overweight} & 185 & (22.8\%) & 158 & (20.9\%) & 343 & (21.9\%) \\
     & \textbf{Obese} & 73 & (8.9\%) & 75 & (9.9\%) & 148 & (9.4\%) \\
     & \textbf{None} & 118 & (14.5\%) & 64 & (8.5\%) & 182 & (11.6\%) \\  
     \midrule
     \multirow{2}{*}{\textbf{Pregnancy status}} & \textbf{True} & 94 & (11.6\%) & 16 & (2.1\%) & 110 & (7.0\%) \\
     & \textbf{False} & 719 & (88.4\%) & 739 & (97.9\%) & 1,458 & (93.0\%) \\
     \bottomrule
   \end{tabular}
 \end{table*}

The weight and height were used to calculate the body mass index (BMI), which was then combined with the age category to generate an age-corrected BMI (acBMI) feature. Age-correction is needed in case of using BMI for children or adolescents. As only the age category (neonate, infant, child, adolescent) was available instead of numerical age in years, the age-corrected BMI categories (underweight, normal weight, overweight, obese, none) were calculated using the manually extracted values from the Brazilian national standard and WHO data~\cite{conde2006body}.  At least one of the necessary values for calculating the age-corrected BMI was missing for 11.6\% of the participants (see Table~~\ref{tab:demographic}). Note that CirCor22 dataset contains missing values for age, height and/or weight variables, which are treated as a new separate category ("None" in the Age and acBMI categorical variables), thus avoiding data loss, and allowing the model to capture the missingness patterns and potentially improve predictions \cite{Vach98}. The categorical variables for gender and pregnancy status were represented using the simple binary encoding, whereas age and age-corrected BMI were encoded with dummy variables with a missing value indicator represented by a zero vector, leading to 10-dimensional feature vector. Details of the study population are presented in Table~\ref{tab:demographic}.

\subsubsection{Evaluation}
\label{subsubsec:evaluation}

The models for heart sound classification were evaluated on the benchmark CirCor22 test dataset used in the PhysioNet 2022 challenge for the task of clinical outcome classification.

We used the official PhysioNet 2022 challenge metric to rank our models in comparison to the challenge participants; i.e., the nonlinear cost function for expert screening (please see the official challenge website \url{https://moody-challenge.physionet.org/2022/} for more details). Additionally, we also provide results for more common evaluation metrics, such as accuracy, F1 score and area under the receiver operating characteristic curve (AUROC).

\section{Experimental results}
\label{sec:results}

This section summarizes the results of quality assessment of the PCG recordings and heart sound classification.

\subsection{Quality assessment of PCG recordings}
\label{subsec:quality_results}

\begin{table*}[t]
   \caption{Performance of the heart sound quality assessment models trained and tested on the HSCT11, Pascal, CinC16 and CDHS datasets.}
   \label{tab:quality_assessment}
   \scriptsize
   \centering
   \begin{tabular}{p{0.7cm} p{1.2cm} p{1.2cm} p{2.1cm} p{1.4cm} p{1.1cm} p{1.1cm}}
     \toprule
      & \textbf{Accuracy} & \textbf{Precision} & \textbf{Recall \hspace{0.5cm} (Sensitivity)} & \textbf{Specificity} & \textbf{F1 score} & \textbf{AUROC}\\
     \midrule
     \multicolumn{7}{c}{\textit{No feature selection (all features used)}} \\
     \midrule
     \textbf{SVM} & 0.922 & 0.944 & 0.950 & 0.843 & 0.947 & 0.972 \\
     \textbf{RF} & 0.926 & 0.944 & 0.956 & 0.841 & 0.950 & 0.981 \\
     \textbf{GB} & \textbf{0.933} & \textbf{0.954} & 0.956 & \textbf{0.870} & \textbf{0.955} & 0.981 \\
     \textbf{VC} & \textbf{0.933} & 0.949 & \textbf{0.960} & 0.855 & \textbf{0.955} & \textbf{0.982} \\
     \midrule
     \multicolumn{7}{c}{\textit{After feature selection (20\% of the most relevant features)}} \\
     \midrule
     \textbf{SVM} & 0.932 & \textbf{0.952} & 0.956 & \textbf{0.865} & 0.954 & 0.978 \\
     \textbf{RF} & 0.932 & 0.948 & \textbf{0.960} & 0.853 & 0.954 & 0.979 \\
     \textbf{GB} & 0.930 & 0.949 & 0.957 & 0.855 & 0.953 & 0.978 \\
     \textbf{VC} & \textbf{0.933} & 0.951 & 0.958 & 0.862 & \textbf{0.955} & \textbf{0.983} \\
     \bottomrule
   \end{tabular}
\end{table*}

The heart sound quality assessment was performed using HSCT11, Pascal, CinC16 and CDHS datasets, and evaluated with SVM, RF, GB and VC machine learning models, as explained in Section~\ref{subsec:quality}. 
The evaluation across multiple classifiers demonstrates consistently strong performance, with all models achieving accuracy above 0.92 and AUROC above 0.97. While all models already performed well using the full feature set, applying feature selection—retaining only the top 20\% most relevant features—led to slightly improved or more stable results across most metrics, particularly in precision, recall (sensitivity) and specificity. This trend suggests that eliminating redundant or less informative features helps to reduce model variance and prevents overfitting, leading to more reliable decision boundaries. In particular, models such as SVM and VC benefited from this dimensionality reduction, showing improved specificity without compromising sensitivity. The obtained results indicate that carefully selected features not only preserve discriminative capacity but also contribute to greater generalization and robustness, which is especially valuable in clinical settings where consistent performance across diverse recording conditions is essential.

\begin{figure}[t]
 \centering
    \begin{center}
        \includegraphics[width=7cm]{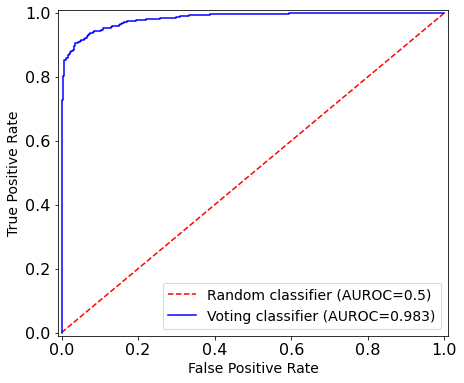}
    \end{center}
    \caption{ROC curve of the best performing model for the heart sound quality assessment (voting classifier).}
    \label{fig:ROC}
\end{figure}

VC with feature selection showed the best overall performance, with an accuracy of 0.933, F1 score of 0.955 and AUROC of 0.983 (see Table~\ref{tab:quality_assessment}). The model had slightly higher sensitivity than specificity, meaning that it was more adapted to identify high quality instances correctly, but may also generate a few false positives, leading to a lower specificity. However, the ROC curve in Figure~\ref{fig:ROC}, plotted at various threshold settings, confirmed a good balance between the sensitivity (true positive rate) and specificity (1 - false positive rate). Therefore, VC after feature selection was used for quality assessments of PCG recordings in all further experiments. Note that quality assessment was considered as a preprocessing step for the subsequent heart sound classification task. Therefore, the emphasis was not on developing state-of-the art deep learning based quality assessment approach, but on getting the reliable estimates of quality labels, which proved to be sufficient with VC quality classifier. Furthermore, the proposed VC classifier is less computationally demanding than the deep learning based approaches.
 
\subsection{Heart sound classification from PCG recordings}
\label{sec:heart_class_results}
We ran two sets of experiments, without and with the heart sound quality assessment, and benchmarked the performance of the proposed models on the PhysioNet 2022 heart sound classification challenge. Experiments without the heart sound quality assessment are meant to provide a fair comparison with the existing results available for the challenge dataset, whereas the quality assessment was used to remove the low quality PCG samples, thus allowing for training and evaluating the models on cleaner data, and assessing the influence of the quality assessment process on model performance. 

\begin{figure*}[htbp]
 \centering
    \begin{center}
        \includegraphics[width=10cm]{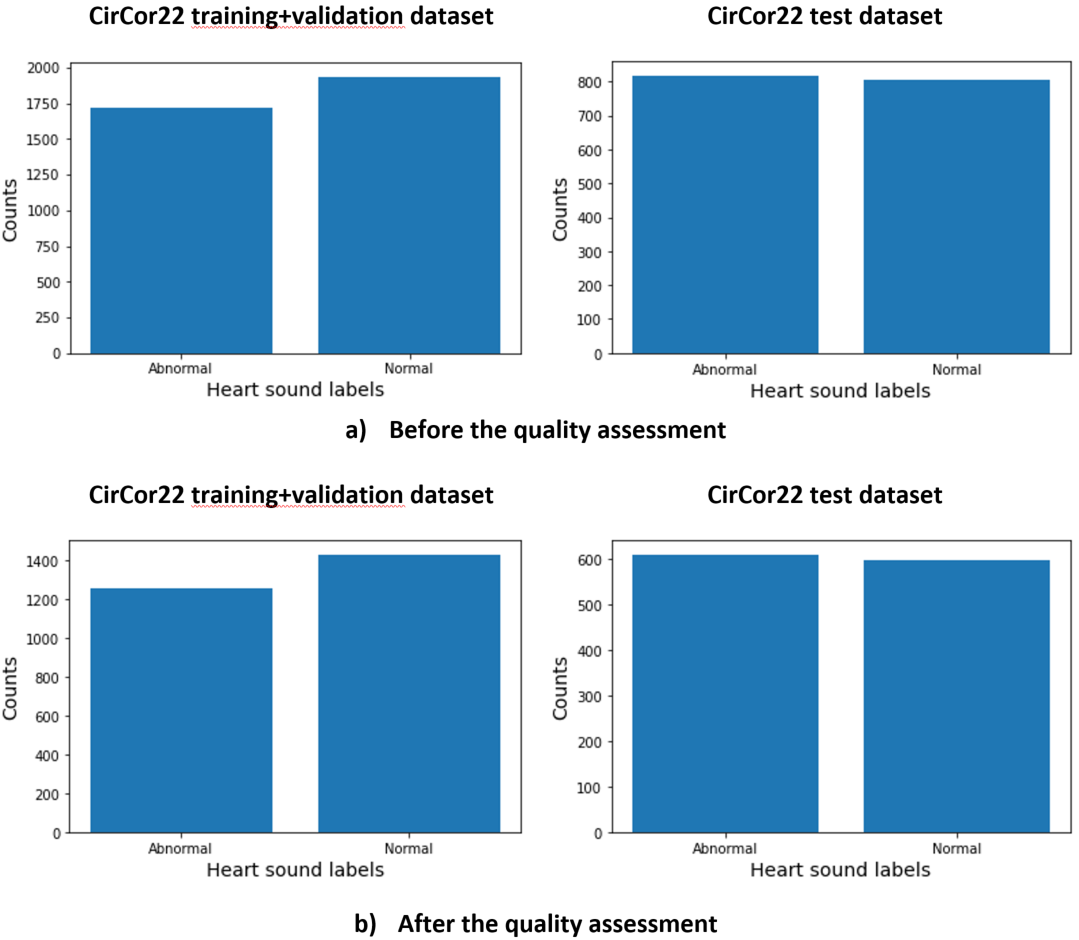}
    \end{center}
    \caption{Distribution of the clinical outcome heart sound labels in the CirCor22 dataset (PhysioNet 2022 challenge): a) Before; and b) After the quality assessment.}
    \label{fig:CirCor22_distribution}
\end{figure*} 

The first set of experiments was done without the heart sound quality assessment. The distribution of clinical outcome labels (normal/abnormal) shown in Figure \ref{fig:CirCor22_distribution}a indicates that both the training and test subsets were rather well balanced. Given that CirCor22 dataset contains, besides PCG recordings, additional socio-demographic information, two analyses were done, namely, unimodal, based only on audio features, and multimodal, that integrated audio with socio-demographic variables (age, sex, acBMI and pregnancy status) using a feature-level fusion.

\begin{table*}[tbp]
   \caption{Performance of the heart sound classification before the quality assessment. The evaluation was done using the CirCor22 test subset (PhysioNet 2022 challenge).}
   \label{tab:PhysioNet22a}
   \scriptsize
   \centering
   \hspace*{-3cm}\begin{tabular}{llccccccc}
     \toprule
      &  & \textbf{Modality} & \textbf{Sampling frequency} & \textbf{Rank} & \textbf{Cost} & \textbf{Accuracy} & \textbf{F1 score} & \textbf{AUROC} \\
    \midrule
      \multirow{5}{*}{\textbf{Challenge}} & CUED\_Acoustics~\cite{mcdonald2022detection} & Audio/dem. & 4 kHz & 1* & 11,144 & 0.602 & 0.549 & 0.693 \\ 
      & prna~\cite{chang2022multi} & Audio/dem. & 4 kHz & 2* & 11,403 & 0.587 & 0.536 & 0.691 \\
      & Melbourne\_Kangas~\cite{imran2022fusion} & Audio/dem. & 4 kHz & 3* & 11,735 & 0.568 & 0.528 & 0.663 \\ 
      & CeZIS~\cite{bruoth2022murmur} & Audio/dem. & 4 kHz & 4* & 11,916 & 0.560 & 0.511 & 0.614 \\
      & CAU\_UMN~\cite{lee2022deep} & Audio/dem. & 4 kHz & 5* & 11,933 & 0.562 & 0.505 & 0.660 \\
      & HCCL~\cite{Kim2022} & Audio/dem. & 4 kHz & 6* & 11,943 & 0.562 & 0.532 & 0.605 \\
      & Listen2YourHeart~\cite{ballas2022listen2yourheart} & Audio & 2 kHz & 7* & 11,946 & 0.558 & 0.512 & 0.627 \\
    \midrule
      \multirow{4}{*}{\textbf{CardioPHON}} & \multirow{4}{*}{Feature extraction} & Audio & 1 kHz & 25 & 13,503 & 0.562 & 0.559 & 0.627 \\    
      & & Audio & 4 kHz & 11 & 12,173 & 0.595 & 0.588 & 0.658 \\                            
      & & Audio/dem. & 1 kHz & 18 & 12,780 & 0.581  & 0.576 & 0.656 \\                     
      & & Audio/dem. & 4 kHz & 13 & 12,327 & 0.600 & 0.595 & 0.665 \\                           
    \midrule
     \multirow{4}{*}{\textbf{CardioPHON}} & \multirow{4}{*}{Training from scratch} & Audio & 1 kHz & 16 & 12,579 & 0.587 & 0.582 & 0.650 \\
      & & Audio & 4 kHz & 4 & 11,887 & 0.585 & 0.571 & 0.677 \\               
      & & Audio/dem. & 1 kHz & 15 & 12,536 & 0.581 & 0.574 & 0.652 \\
      & & Audio/dem. & 4 kHz & \textbf{1} & \textbf{11,107} & \textbf{0.625} & \textbf{0.612} & \textbf{0.693} \\
    \midrule
     \multirow{4}{*}{\textbf{CardioPHON}} & \multirow{4}{*}{Fine-tuning} & Audio & 1 kHz & 13 & 12,342 & 0.570 & 0.558 & 0.656 \\
      & & Audio & 4 kHz & 4 & 11,788 & 0.593 & 0.581 & 0.663 \\                    
      & & Audio/dem. & 1 kHz & 11 & 12,125 & 0.587 & 0.577 & 0.676 \\
      & & Audio/dem. & 4 kHz & 3 & 11,622 & 0.612 & 0.603 & 0.683 \\
    \bottomrule
   \end{tabular}
   \vskip 3mm   
    *The official ranking of the PhysioNet 2022 challenge. Our CardioPHON models are ranked with respect to the position on the official leaderboard.\\
\end{table*}

\begin{table*}[htbp]
   \caption{Performance of the heart sound classification after the quality assessment. The evaluation was done using the CirCor22 test subset (PhysioNet 2022 challenge).}
   \label{tab:PhysioNet22b}
   \scriptsize
   \centering
   \hspace*{-3cm}\begin{tabular}{llccccccc}
     \toprule
      &  & \textbf{Modality} & \textbf{Sampling frequency} & \textbf{Rank} & \textbf{Cost} & \textbf{Accuracy} & \textbf{F1 score} & \textbf{AUROC} \\
    \midrule
      \multirow{5}{*}{\textbf{Challenge}} & CUED\_Acoustics~\cite{mcdonald2022detection} & Audio/dem. & 4 kHz & 1* & 11,144 & 0.602 & 0.549 & 0.693 \\ 
      & prna~\cite{chang2022multi} & Audio/dem. & 4 kHz & 2* & 11,403 & 0.587 & 0.536 & 0.691 \\
      & Melbourne\_Kangas~\cite{imran2022fusion} & Audio/dem. & 4 kHz & 3* & 11,735 & 0.568 & 0.528 & 0.663 \\ 
      & CeZIS~\cite{bruoth2022murmur} &  Audio/dem. & 4 kHz & 4* & 11,916 & 0.560 & 0.511 & 0.614 \\
      & CAU\_UMN~\cite{lee2022deep} & Audio/dem. & 4 kHz & 5* & 11,933 & 0.562 & 0.505 & 0.660 \\
      & HCCL~\cite{Kim2022} & Audio/dem. & 4 kHz & 6* & 11,943 &0.562 & 0.532 & 0.605 \\
      & Listen2YourHeart~\cite{ballas2022listen2yourheart} & Audio & 2 kHz & 7* & 11,946 & 0.558 & 0.512 & 0.627 \\
    \midrule
      \multirow{2}{*}{\textbf{CardioPHON}} & \multirow{2}{*}{Feature extraction} & Audio & 1 kHz & 14 & 12,415 & 0.589 & 0.583 & 0.638 \\ 
      & & Audio/dem. & 1 kHz & 14 & 12,377 & 0.600 & 0.595 & 0.656 \\
    \midrule
     \multirow{2}{*}{\textbf{CardioPHON}} & \multirow{2}{*}{Training from scratch} & Audio & 1 kHz & 18 & 12,712 & 0.562 & 0.553 & 0.630 \\
      & & Audio/dem. & 1 kHz & 11 & 12,124 & 0.577 & 0.563 & 0.646 \\ 
    \midrule
     \multirow{2}{*}{\textbf{CardioPHON}} & \multirow{2}{*}{Fine-tuning} & Audio & 1 kHz & 11 & 12,113 & 0.574 & 0.560 & 0.651 \\      
     & & Audio/dem. & 1 kHz & 4 & 11,877 & 0.595 & 0.585 & 0.664 \\                                  
     \bottomrule
   \end{tabular}
   \vskip 3mm   
    *The official ranking of the PhysioNet 2022 challenge. Our CardioPHON models are ranked with respect to the position on the official leaderboard.\\
\end{table*}

\begin{table}[h]
   \caption{Per-class performance of the best CardioPHON model for heart sound classification. The evaluation was done using the CirCor22 test subset (PhysioNet 2022 challenge).}
   \label{tab:per-class-performance}
   \footnotesize
   \centering
   \begin{tabular}{lcc}
     \toprule
      & \textbf{Abnormal} & \textbf{Normal} \\
     \midrule
     \textbf{Accuracy} & 0.793 & 0.451 \\
     \textbf{F1 score} & 0.682 & 0.542 \\
     \bottomrule
   \end{tabular}
\end{table}

Experimental results for three setups (feature extraction, training from scratch and fine-tuning) using unimodal (audio) and multimodal features (audio/socio-demographic), without the quality assessment, are presented in Table~\ref{tab:PhysioNet22a}. We provide results for two sampling frequencies, i.e, the native sampling frequency of CirCor22 dataset (4kHz), as well as 1 kHz, which corresponds to the sampling frequency of datasets used for developing the quality assessment model. Furthermore we show, for comparison, the results obtained for the seven best performing teams of the PhysioNet 2022 challenge, with the information on modality type that was deployed. It is interesting to note here that first six of these models were using audio combined with the socio-demographic data in the multimodal setup. Only the seventh placed model was using audio exclusively.

The best performing CardioPHON model was trained from scratch in an SSL fashion on the collection of PCG recordings from six publicly available datasets (Pascal, HSCT11, CDHS, Ephnogram, Open Heart and CinC16), sampled at frequency of 4 kHz, and used the combination of audio and socio-demographic features. It achieved the best ranking on the official challenge leaderboard, outperforming all challenge participants, with the nonlinear cost for expert screening equal to 11,107, as well as the best accuracy (0.625), F1 score (0.612) and AUROC (0.693).

After quality assessment, more than 25\% of the recordings were removed, as their quality was considered unacceptable (see Figure~\ref{fig:CirCor22_distribution}b). The performance of almost all models had improved for the same sampling frequency of 1 kHz, as shown in Table~\ref{tab:PhysioNet22b}, confirming that quality assessment might be an important preprocessing step to guarantee reliable evaluation of clinical outcomes. 

The performance of the best CardioPHON model (fused audio and socio-demographic features, 4 kHz sampling frequency, model trained from scratch, before the quality assessment) per each class separately is shown in Table~\ref{tab:per-class-performance}. The model performed better for the "Abnormal" than for the "Normal" class, with substantially higher accuracy and F1 score.

\section{Discussion}
\label{sec:discussion}
\subsection{Discussion of quality assessment of PCG recordings}
\label{subsec:quality_discussion}

Capturing heart sounds, either by digital stethoscopes or by mobile phones without the external noise, is difficult to achieve due to physiological body noises, as well as ambient artifacts, both in clinical and non-clinical settings \cite{SHARIATPANAH23}. Noise contamination in PCG recordings can harm feature extraction, making the features less representative of the underlying audio content; thus, hindering the model's ability to learn relevant patterns. As a consequence, this could mislead the model during training, reducing its performance \cite{Jain17}. Thus, identifying low-quality PCG recordings is critical to ensure the consistency and reliability of the models for screening of cardiac function. 

We developed a model for quality assessment of PCG recordings trained on the annotated data containing 8,359 recordings from four datasets (HSCT11, Pascal, CinC16 and CDHS). We used the trained model to generate the pseudo quality labels for the remaining non-annotated datasets (Ephnogram, Open Heart and CirCor22), and further as the preprocessing step for the classification of heart sounds, to identify and remove PCG recordings that did not meet the minimal quality requirements for further processing. 

To understand the model's behavior and provide insights into its strengths and weaknesses better, we analyzed the feature importance for PCG quality assessment in Figure~\ref{fig:feat_importance}. We found that S2 HSMM quality factor, which is determined as the mean height of S2 envelope divided by the average of the mean height of diastole and systole \cite{Shi20}, is the most informative feature. This suggests that the second heart sound (S2), which occurs at the end of ventricular systole and beginning of ventricular diastole, might carry more relevant information for capturing signal quality than the first heart sound (S1). Power spectrum features have also experienced high feature importance, with the ability to highlight the presence of noise in PCG recordings. 

\subsection{Discussion of heart sound classification from PCG recordings}
\label{sec:heart_class_discussion}

The results obtained for heart sound classification from PCG recordings highlight a clear advantage of domain adaptation when using SSL approaches. Specifically, both training a model from scratch using SSL on PCG recordings and fine-tuning a model pretrained on general audio significantly outperform the feature extraction (zero-shot learning) approach, where the pretrained model is used without further adaptation. This performance gap underscores the importance of aligning the model’s internal representations with the specific characteristics of heart sounds, which differ substantially from general audio data in terms of frequency content, temporal structure, and clinical relevance. Notably, training from scratch enables the model to learn representations that are fully tailored to the PCG domain, capturing fine-grained diagnostic features such as subtle variations in murmur intensity, timing, and rhythm that may be absent or underrepresented in general audio corpora. On the other hand, fine-tuning leverages general audio pretraining to provide a strong initialization, allowing the model to converge more quickly and potentially generalize better, especially when labeled PCG data is limited. The comparable performance of both strategies suggests that while general audio features offer a useful foundation, targeted adaptation to the heart sound domain is critical to achieving optimal classification results. In contrast, the feature extraction (zero-shot learning) approach suffers from a domain mismatch between general-purpose audio and PCG recordings. Without adaptation, the model’s learned features are not sufficiently aligned with the physiological patterns and pathological cues present in heart sounds, resulting in degraded performance. This finding reinforces the necessity of domain-aware representation learning in medical audio tasks, where even subtle mismatches between source and target domains can lead to clinically significant drops in performance.

Moreover, SSL helps the model to learn invariant and more robust feature representations, making it more adaptable to variations across datasets recorded in different conditions. This leads to improved performance on downstream tasks such as heart sound classification. There is only one reported attempt of using SSL approach for PCG signal \cite{ballas2022listen2yourheart}; however, this model is pretrained using only two datasets coming from the  PhysioNet/Computing in Cardiology Challenge 2016 and the George B. Moody PhysioNet Challenge  2022. We aggregate all publicly available PCG datasets for training, including diverse patient populations, clinical and non-clinical settings, and a broad range of cardiac conditions, enabling the model to learn representations that generalize across devices, acquisition protocols, and patient demographics. Another important difference is in the model architecture: Model used in \cite{ballas2022listen2yourheart} relies on SimCLR, a contrastive learning method that learns representations by bringing positive pairs (augmented views of the same sample) closer together and pushing negative pairs (different samples) apart in the embedding space. We rely on BYOL-A, a non-contrastive learning method that learns by predicting one augmented view of a sample from another, thereby requiring no negative pairs. This provides more stable and data-efficient training, especially in domains where distinguishing negatives is difficult. This is a critical advantage in the domain of heart sounds, where clinically similar heart sounds, potentially from patients with the same or related conditions, could mistakenly be treated as negatives, degrading representation quality and downstream task performance.

Using the native sampling frequency of 4 kHz for the PhysioNet 2022 challenge improves the performance for all setups substantially, i.e., the downsampling introduced an irreversible loss of information. 

Multimodal CardioPHON model, with fused audio and socio-demographic features, tended to be superior over the unimodal (audio only) for all setups and over all sampling rates, i.e., the socio-demographic variables were increasing the predictive capacity of audio features, uncovering the full potential of multimodal data fusion. CardioPHON model that used the fused audio and socio-demographic features achieved state-of-the-art results, with the best ranking on the official challenge leaderboard. Our best unimodal CardioPHON model, that utilized only the audio modality, was also ranked the first among unimodal models (total rank 4, see Table~\ref{tab:PhysioNet22a}), leaving behind the models that made leverage of the multiple information sources. 

By analyzing per class performance of the best multimodal CardioPHON model in Table~\ref{tab:per-class-performance}, one can conclude that the model performed substantially better for the "Abnormal" than for the "Normal" clinical outcome. This behavior may be desirable for pre-screening abnormal cardiac function, as the model was able to identify most of the patients with abnormal heart sounds correctly, even if it assumed more false positive results. This is especially important for identifying cardiac conditions that require immediate and timely intervention, where missing true positive cases may lead to a fatal outcome.

It might be interesting to note here that three out of seven best performing challenge approaches, i.e., CUED\_Acoustics, Melbourne\_Kangas and HCCL, integrate heart sound segmentation (HSS) with heart sound classification, whereas the proposed method was able to outperform them without taking HSS into account.

Quality assessment of PCG recordings improved the performance for almost all the setups, assuming that the sampling frequency of 1 kHz was used. However, due to a clear mismatch between the sampling frequencies used in the PhysioNet 2022 challenge (4 kHz), and the model for quality assessment of PCG recordings (1 kHz), an inevitable information loss occurred which resulted in an inferior performance of the classification models developed for sampling frequency of 1 kHz in comparison to the models designed for sampling frequency of 4 kHz. Despite the improvement that quality assessment process brought, the best achieved ranking has dropped to the fourth place on the official leaderboard. We believe that a quality assessment model developed for a sampling frequency of 4 kHz would unlock the full potential of the quality assessment process and even further increase the performance of the proposed approach.

The main limitation of the study is related to the PCG quality assessment model, which was developed for the sampling frequency of 1 kHz only. The choice was driven by the sampling frequency of the datasets used for model design. Given the fact that the frequency content of S1 and S2 sounds is mostly in the range of 20-500 Hz \cite{DEBBAL2007}, a sampling frequency of 1 kHz should, in theory, be sufficient for analyzing PCG recordings, as it is supposed to capture main frequency components of the heart sounds. However, this may still be a limiting factor for certain applications, as shown in our experiments.

\section{Conclusion}
\label{sec:conclusion}

In this work, we proposed the novel heart sound quality assessment model, and integrated it with the heart sound classification model trained in the self-supervised learning manner. We show that self-supervised pretraining on multiple small- and mid-sized datasets is an essential step for building a robust heart sound classification model, capturing wide range of heart sound variations across different datasets. The multimodal CardioPHON model, that uses the combination of audio and socio-demographic features, achieved the best ranking on the official challenge leaderboard of the PhysioNet 2022 challenge, proving that different modalities carry distinct, but complementary information, leading to a more comprehensive and better feature representation.
Moreover, the best unimodal CardioPHON model, utilizing only the audio modality, was ranked first among the unimodal models (total rank 4), leaving behind the models deploying the multiple information sources. 

Future work will integrate heart sound segmentation with heart sound classification, for potentially improved performance. Moreover, given that there is still a lack of large-scale heart sound data, audio data augmentation will be considered to increase data diversity and robustness to noise additionally; thus, improving the generalization and performance of the heart sound classification models.

\section{Data availability}

The pretrained CardioPHON model, heart sound quality labels for the publicly available heart sound datasets and the source code are available at \url{https://git.lih.lu/vdespotovic/cardiophon}.

\section{Acknowledgements}
This publication is based upon work from COST Action CA19121 -- GoodBrother, Network on Privacy-Aware Audio- and Video-Based Applications for Active and Assisted Living (\url{https://goodbrother.eu/}), supported by COST (European Cooperation in Science and Technology) (\url{https://www.cost.eu/}). Andrej Zgank's research work was partially supported by the Slovenian Research and Innovation Agency (research core funding No. P2-0069 Advanced Methods of Interaction in Telecommunications). 

We would like to thank Jorge Oliveira (Universidade Portucalense Infante D. Henrique, Portugal) and Hong Tang (Dalian University of Technology, China) for providing us with the validation and test subset of the CirCor22 dataset, and the valuable insights when it comes to the CDHS dataset, respectively.

\section{Role of the funding source}
The study funders had no role in the study design; in the collection, analysis, and interpretation of data; in the writing of the report; and in the decision to submit the article for publication.

\section{Author contributions}

\textbf{V.D., P.P. and A. Z.} have contributed to the conceptualization, methodology, software development, data curation, drafting the article, visualization, and final approval of the article.

\bibliography{mybib}

@inproceedings{mcdonald2022detection,
  title={Detection of Heart Murmurs in Phonocardiograms with Parallel Hidden Semi-Markov Models},
  author={McDonald, Andrew and Gales, Mark JF and Agarwal, Anurag},
  booktitle={2022 Computing in Cardiology (CinC)},
  volume={498},
  pages={1--4},
  year={2022},
  organization={IEEE}
}

@inproceedings{bruoth2022murmur,
  title={Murmur Identification Using Supervised Contrastive Learning},
  author={Bruoth, Erik and Bugata, Peter and Gajdo{\v{s}}, D{\'a}vid and Hud{\'a}k, D{\'a}vid and Kme{\v{c}}ov{\'a}, Vladim{\'\i}ra and Sta{\v{n}}kov{\'a}, Monika and Szabari, Alexander and Voz{\'a}rikov{\'a}, Gabriela and others},
  booktitle={2022 Computing in Cardiology (CinC)},
  volume={498},
  pages={1--4},
  year={2022},
  organization={IEEE}
}

@inproceedings{lee2022deep,
  title={Deep Learning Based Heart Murmur Detection Using Frequency-time Domain Features of Heartbeat Sounds},
  author={Lee, Jungguk and Kang, Taein and Kim, Narin and Han, Soyul and Won, Hyejin and Gong, Wuming and Kwak, Il-Youp},
  booktitle={2022 Computing in Cardiology (CinC)},
  volume={498},
  pages={1--4},
  year={2022}
}

@inproceedings{chang2022multi,
  title={Multi-Task Prediction of Murmur and Outcome from Heart Sound Recordings},
  author={Chang, Yale and Liu, Luoluo and Antonescu, Corneliu},
  booktitle={2022 Computing in Cardiology (CinC)},
  volume={498},
  pages={1--4},
  year={2022}
}

@inproceedings{imran2022fusion,
  title={A Fusion of Handcrafted Feature-Based and Deep Learning Classifiers for Heart Murmur Detection},
  author={Imran, Zaria and Grooby, Ethan and Malgi, Vinayaka Vivekananda and Sitaula, Chiranjibi and Aryal, Sunil and Marzbanrad, Faezeh},
  booktitle={2022 Computing in Cardiology (CinC)},
  volume={498},
  pages={1--4},
  year={2022}
}

@article{springer2016automated,
  title={Automated signal quality assessment of mobile phone-recorded heart sound signals},
  author={Springer, David B and Brennan, Thomas and Ntusi, Ntobeko and Abdelrahman, Hassan Y and Z{\"u}hlke, Liesl J and Mayosi, Bongani M and Tarassenko, Lionel and Clifford, Gari D},
  journal={Journal of medical engineering \& technology},
  volume={40},
  number={7-8},
  pages={342--355},
  year={2016}
}

@inproceedings{zabihi2016heart,
  title={Heart sound anomaly and quality detection using ensemble of neural networks without segmentation},
  author={Zabihi, Morteza and Rad, Ali Bahrami and Kiranyaz, Serkan and Gabbouj, Moncef and Katsaggelos, Aggelos K},
  booktitle={2016 Computing in Cardiology Conference (CinC)},
  pages={613--616},
  year={2016}
}

@inproceedings{das2017novel,
  title={Novel features from autocorrelation and spectrum to classify phonocardiogram quality},
  author={Das, Deepan and Banerjee, Rohan and Choudhury, Anirban Dutta and Bhattacharya, Sakyajit and Deshpande, Parijat and Pal, Arpan and Mandana, Kayapanda M},
  booktitle={2017 39th Annual International Conference of the IEEE Engineering in Medicine and Biology Society (EMBC)},
  pages={4516--4520},
  year={2017}
}

@article{akram2018analysis,
  title={Analysis of {PCG} signals using quality assessment and homomorphic filters for localization and classification of heart sounds},
  author={Akram, Muhammad Usman and Shaukat, Arslan and Hussain, Farhan and Khawaja, Sajid Gul and Butt, Wasi Haider and others},
  journal={Computer methods and programs in biomedicine},
  volume={164},
  pages={143--157},
  year={2018}
}

@article{tang2021automated,
  title={Automated signal quality assessment for heart sound signal by novel features and evaluation in open public datasets},
  author={Tang, Hong and Wang, Miao and Hu, Yating and Guo, Binbin and Li, Ting},
  journal={BioMed Research International},
  volume={2021},
  year={2021}
}

@article{mei2021classification,
  title={Classification of heart sounds based on quality assessment and wavelet scattering transform},
  author={Mei, Na and Wang, Hongxia and Zhang, Yatao and Liu, Feifei and Jiang, Xinge and Wei, Shoushui},
  journal={Computers in Biology and Medicine},
  volume={137},
  pages={104814},
  year={2021}
}

@article{grooby2021neonatal,
  title={Neonatal heart and lung sound quality assessment for robust heart and breathing rate estimation for telehealth applications},
  author={Grooby, Ethan and He, Jinyuan and Kiewsky, Julie and Fattahi, Davood and Zhou, Lindsay and King, Arrabella and Ramanathan, Ashwin and Malhotra, Atul and Dumont, Guy A and Marzbanrad, Faezeh},
  journal={IEEE Journal of Biomedical and Health Informatics},
  volume={25},
  number={12},
  pages={4255--4266},
  year={2021}
}

@article{grooby2022real,
  title={Real-time multi-level neonatal heart and lung sound quality assessment for telehealth applications},
  author={Grooby, Ethan and Sitaula, Chiranjibi and Fattahi, Davood and Sameni, Reza and Tan, Kenneth and Zhou, Lindsay and King, Arrabella and Ramanathan, Ashwin and Malhotra, Atul and Dumont, Guy Albert and others},
  journal={IEEE Access},
  volume={10},
  pages={10934--10948},
  year={2022}
}

@article{despotovic2022audio,
  title={Audio-based Active and Assisted Living: A review of selected applications and future trends},
  author={Despotovic, Vladimir and Pocta, Peter and Zgank, Andrej},
  journal={Computers in Biology and Medicine},
  pages={106027},
  year={2022}
}

@inproceedings{reyna2022heart,
  title={Heart murmur detection from phonocardiogram recordings: The {G}eorge {B}. {M}oody {P}hysionet challenge 2022},
  author={Reyna, Matthew A and Kiarashi, Yashar and Elola, Andoni and Oliveira, Jorge and Renna, Francesco and Gu, Annie and Alday, Erick A Perez and Sadr, Nadi and Sharma, Ashish and Mattos, Sandra and others},
  booktitle={2022 Computing in Cardiology (CinC)},
  volume={498},
  pages={1--4},
  year={2022}
}

@article{clifford2017recent,
  title={Recent advances in heart sound analysis},
  author={Clifford, Gari D and Liu, Chengyu and Moody, Benjamin E and Roig, Jos{\'e} Millet and Schmidt, Samuel E and Li, Qiao and Silva, Ikaro and Mark, Roger G},
  journal={Physiological measurement},
  volume={38},
  pages={E10--E25},
  year={2017}
}

@article{chen2021deep,
  title={Deep learning methods for heart sounds classification: a systematic review},
  author={Chen, Wei and Sun, Qiang and Chen, Xiaomin and Xie, Gangcai and Wu, Huiqun and Xu, Chen},
  journal={Entropy},
  volume={23},
  number={6},
  pages={667},
  year={2021}
}

@article{chan2019contactless,
  title={Contactless cardiac arrest detection using smart devices},
  author={Chan, Justin and Rea, Thomas and Gollakota, Shyamnath and Sunshine, Jacob E},
  journal={NPJ digital medicine},
  volume={2},
  number={1},
  pages={52},
  year={2019}
}

@article{Ross2014,
    author = {Ross, Brian C.},
    journal = {PLOS ONE},
    publisher = {Public Library of Science},
    title = {Mutual Information between Discrete and Continuous Data Sets},
    year = {2014},
    volume = {9},
    pages = {1-5},
    number = {2}
}

@inproceedings{Gemmeke2017,
title	= {Audio {S}et: An ontology and human-labeled dataset for audio events},
author	= {Jort F. Gemmeke and Daniel P. W. Ellis and Dylan Freedman and Aren Jansen and Wade Lawrence and R. Channing Moore and Manoj Plakal and Marvin Ritter},
year	= {2017},
booktitle	= {Proc. IEEE ICASSP 2017}
}

@inproceedings{niizumi2021byol-a,
    title={{BYOL for Audio}: Self-Supervised Learning for General-Purpose Audio Representation},
    author={Daisuke Niizumi and Daiki Takeuchi and Yasunori Ohishi and Noboru Harada and Kunio Kashino},
    booktitle = {2021 International Joint Conference on Neural Networks (IJCNN)},
    year={2021}
}

@article{niizumi2023byol-a,
    title={{BYOL for Audio}: Exploring Pre-trained General-purpose Audio Representations},
    author={Niizumi, Daisuke and Takeuchi, Daiki and Ohishi, Yasunori and Harada, Noboru and Kashino, Kunio},
    journal={IEEE/ACM Transactions on Audio, Speech, and Language Processing}, 
    publisher={Institute of Electrical and Electronics Engineers (IEEE)},
    year={2023},
    volume={31},
    pages={137–-151}
}

@ARTICLE{Shi20,
  author={Shi, Kilin and Schellenberger, Sven and Michler, Fabian and Steigleder, Tobias and Malessa, Anke and Lurz, Fabian and Ostgathe, Christoph and Weigel, Robert and Koelpin, Alexander},
  journal={IEEE Transactions on Biomedical Engineering}, 
  title={Automatic Signal Quality Index Determination of Radar-Recorded Heart Sound Signals Using Ensemble Classification}, 
  year={2020},
  volume={67},
  number={3},
  pages={773--785}
}

@incollection{Vach98,
  title={Missing data in epidemiological studies},
  author={Vach, W. and Blettner, M.},
  booktitle={Encyclopedia of Biostatistics},
  editor={Armitrage, P. and Colton, T.},
  pages={2641--2654},
  year={1998}
}

@inproceedings{Grill20,
author = {Grill, Jean-Bastien and Strub, Florian and Altch\'{e}, Florent and Tallec, Corentin and Richemond, Pierre H. and Buchatskaya, Elena and Doersch, Carl and Pires, Bernardo Avila and Guo, Zhaohan Daniel and Azar, Mohammad Gheshlaghi and Piot, Bilal and Kavukcuoglu, Koray and Munos, R\'{e}mi and Valko, Michal},
title = {Bootstrap Your Own Latent a New Approach to Self-Supervised Learning},
year = {2020},
booktitle = {Proceedings of the 34th International Conference on Neural Information Processing Systems},
articleno = {1786},
numpages = {14},
series = {NIPS'20}
}

@article{SHARIATPANAH23,
title = {Exploring the impact of noise and degradations on heart sound classification models},
journal = {Biomedical Signal Processing and Control},
volume = {85},
pages = {104932},
year = {2023},
author = {Davoud {Shariat Panah} and Andrew Hines and Susan McKeever}
}

@article{Jain17,
  title={An adaptive thresholding method for the wavelet based denoising of phonocardiogram signal},
  author={Puneet Kumar Jain and Anil Kumar Tiwari},
  journal={Biomedical Signal Processing and Control},
  year={2017},
  volume={38},
  pages={388-399}
}

@article{DEBBAL2007,
title = {Time-frequency analysis of the first and the second heartbeat sounds},
journal = {Applied Mathematics and Computation},
volume = {184},
number = {2},
pages = {1041-1052},
year = {2007},
author = {S.M. Debbal and F. Bereksi-Reguig}
}

@inproceedings{potes2016ensemble,
  title={Ensemble of feature-based and deep learning-based classifiers for detection of abnormal heart sounds},
  author={Potes, Cristhian and Parvaneh, Saman and Rahman, Asif and Conroy, Bryan},
  booktitle={2016 Computing in Cardiology Conference (CinC)},
  pages={621--624},
  year={2016}
}

@article{Lin21,
title = {Wearable sensors and devices for real-time cardiovascular disease monitoring},
journal = {Cell Reports Physical Science},
volume = {2},
number = {8},
pages = {100541},
year = {2021},
author = {Jian Lin and Rumin Fu and Xinxiang Zhong and Peng Yu and Guoxin Tan and Wei Li and Huan Zhang and Yangfan Li and Lei Zhou and Chengyun Ning}
}

@inproceedings{radford2023robust,
  title={Robust speech recognition via large-scale weak supervision},
  author={Radford, Alec and Kim, Jong Wook and Xu, Tao and Brockman, Greg and McLeavey, Christine and Sutskever, Ilya},
  booktitle={International Conference on Machine Learning},
  pages={28492--28518},
  year={2023},
  organization={PMLR}
}

@inproceedings{Baevski20,
author = {Baevski, Alexei and Zhou, Henry and Mohamed, Abdelrahman and Auli, Michael},
title = {Wav2vec 2.0: A Framework for Self-Supervised Learning of Speech Representations},
year = {2020},
booktitle = {Proceedings of the 34th International Conference on Neural Information Processing Systems},
articleno = {1044},
series = {NIPS'20}
}

@ARTICLE{Oliveira22,
  author={Oliveira, Jorge and Renna, Francesco and Costa, Paulo Dias and Nogueira, Marcelo and Oliveira, Cristina and Ferreira, Carlos and Jorge, Alípio and Mattos, Sandra and Hatem, Thamine and Tavares, Thiago and Elola, Andoni and Rad, Ali Bahrami and Sameni, Reza and Clifford, Gari D. and Coimbra, Miguel T.},
  journal={IEEE Journal of Biomedical and Health Informatics}, 
  title={The {CirCor DigiScope} Dataset: From Murmur Detection to Murmur Classification}, 
  year={2022},
  volume={26},
  number={6},
  pages={2524-2535}
}

@article{Liu2016,
  title={An open access database for the evaluation of heart sound algorithms},
  author={Chengyu Liu and David B. Springer and Qiao Li and Benjamin Moody and Ricardo Abad Juan and Francisco Javier Chorro and Francisco Castells and Jos{\'e} Millet Roig and Ikaro Silva and Alistair E. W. Johnson and Zeeshan Syed and Samuel Emil Schmidt and Chrysa D. Papadaniil and Leontios J. Hadjileontiadis and Hosein Naseri and Ali Moukadem and Alain Dieterlen and Christian Brandt and Hong Tang and Maryam Samieinasab and Mohammad Reza Samieinasab and Reza Sameni and Roger G. Mark and Gari D. Clifford},
  journal={Physiological Measurement},
  year={2016},
  volume={37},
  pages={2181 - 2213}
}

@inproceedings{Gomes2013,
  title={Classifying Heart Sounds - Approaches to the {PASCAL} Challenge},
  author={Elsa Ferreira Gomes and Peter John Bentley and Emanuel Pereira and Miguel Tavares Coimbra and Yiqi Deng},
  booktitle={International Conference on Health Informatics},
  year={2013}
}

@INPROCEEDINGS{Spadaccini13,
  author={Spadaccini, Andrea and Beritelli, Francesco},
  booktitle={2013 18th International Conference on Digital Signal Processing (DSP)}, 
  title={Performance evaluation of heart sounds biometric systems on an open dataset}, 
  year={2013},
  volume={},
  number={},
  pages={1-5}
}

@article{yaseen2018classification,
  title={Classification of heart sound signal using multiple features},
  author={Yaseen and Son, Gui-Young and Kwon, Soonil},
  journal={Applied Sciences},
  volume={8},
  number={12},
  pages={2344},
  year={2018}
}

@article{williams2012standard,
  title={Standard 6: Age groups for pediatric trials},
  author={Williams, Katrina and Thomson, Denise and Seto, Iva and Contopoulos-Ioannidis, Despina G and Ioannidis, John PA and Curtis, Sarah and Constantin, Evelyn and Batmanabane, Gitanjali and Hartling, Lisa and Klassen, Terry},
  journal={Pediatrics},
  volume={129},
  number={Supplement\_3},
  pages={S153--S160},
  year={2012}
}

@article{conde2006body,
  title={Body mass index cutoff points for evaluation of nutritional status in {Brazilian} children and adolescents},
  author={Conde, Wolney L and Monteiro, Carlos A},
  journal={Jornal de pediatria},
  volume={82},
  pages={266--272},
  year={2006}
}

@inproceedings{ballas2022listen2yourheart,
  title={Listen2yourheart: A self-supervised approach for detecting murmur in heart-beat sounds},
  author={Ballas, Aristotelis and Papapanagiotou, Vasileios and Delopoulos, Anastasios and Diou, Christos},
  booktitle={2022 Computing in Cardiology (CinC)},
  volume={498},
  pages={1--4},
  year={2022}
}

@inproceedings{panah2023exploring,
  title={Exploring wav2vec 2.0 model for heart murmur detection},
  author={Panah, Davoud Shariat and Hines, Andrew and McKeever, Susan},
  booktitle={2023 31st European Signal Processing Conference (EUSIPCO)},
  pages={1010--1014},
  year={2023},
  organization={IEEE}
}

@inproceedings{kamson2024exploring,
  title={Exploring Wav2vec 2.0 Model for Heart Sound Analysis},
  author={Kamson, Alex Paul and Sawant, Akshay V and Ghosh, Prasanta Kumar and Jeevannavar, Satish S},
  booktitle={2024 46th Annual International Conference of the IEEE Engineering in Medicine and Biology Society (EMBC)},
  pages={1--5},
  year={2024},
  organization={IEEE}
}

@inproceedings{niizumi2024exploring,
  title={Exploring pre-trained general-purpose audio representations for heart murmur detection},
  author={Niizumi, Daisuke and Takeuchi, Daiki and Ohishi, Yasunori and Harada, Noboru and Kashino, Kunio},
  booktitle={2024 46th Annual International Conference of the IEEE Engineering in Medicine and Biology Society (EMBC)},
  pages={1--4},
  year={2024},
  organization={IEEE}
}

@article{Mangione97,
    author = {Mangione, Salvatore and Nieman, Linda Z.},
    title = "{Cardiac Auscultatory Skills of Internal Medicine and Family Practice Trainees: A Comparison of Diagnostic Proficiency}",
    journal = {JAMA},
    volume = {278},
    number = {9},
    pages = {717-722},
    year = {1997}
}

@article{mcgee2018auscultation,
  title={Auscultation of the heart: General principles},
  author={McGee, Steven},
  journal={Evidence-based Physical Diagnosis},
  pages={327--332},
  year={2018}
}

@article {Kazemnejad2021,
	author = {Arsalan Kazemnejad and Peiman Gordany and Reza Sameni},
	title = {An Open{\textendash}Access Simultaneous Electrocardiogram and Phonocardiogram Database},
	year = {2021},
	doi = {10.1101/2021.05.17.444563},
	publisher = {PhysioNet},
	journal = {bioRxiv}
}

@article{Liu2022,
title = {Audio self-supervised learning: A survey},
journal = {Patterns},
volume = {3},
number = {12},
pages = {100616},
year = {2022},
doi = {https://doi.org/10.1016/j.patter.2022.100616},
author = {Shuo Liu and Adria Mallol-Ragolta and Emilia Parada-Cabaleiro and Kun Qian and Xin Jing and Alexander Kathan and Bin Hu and Björn W. Schuller}
}

@article{Altuve2020,
title = {Fundamental heart sounds analysis using improved complete ensemble EMD with adaptive noise},
journal = {Biocybernetics and Biomedical Engineering},
volume = {40},
number = {1},
pages = {426-439},
year = {2020},
author = {Miguel Altuve and Luis Suárez and Jeyson Ardila}
}

@inproceedings{Kim2022,
  title={Classification of Phonocardiogram Recordings Using Vision Transformer
Architecture},
  author={Kim, Joonyeob and Park, Gibeom and Sun, Bongwon},
  booktitle={2022 Computing in Cardiology (CinC)},
  volume={498},
  pages={1--4},
  year={2022}
}

\end{document}